\documentclass{article}
\usepackage{spconf,amsmath,graphicx,hyperref}
\usepackage{cite}
\usepackage{amssymb,amsfonts}
\usepackage{algorithmic}
\usepackage{textcomp}
\usepackage{pgfplots}
\usepackage{pgf-pie}
\usepackage{xcolor}
\usepackage{booktabs}
\usepackage{multirow}
\usepackage{makecell}

\title{TAU: A Benchmark for Cultural Sound Understanding Beyond Semantics}
\name{\makecell[c]{Yi-Cheng Lin$^1$, Yu-Hua Chen$^2$, Jia-Kai Dong$^1$, Yueh-Hsuan Huang$^1$, Szu-Chi Chen$^1$, \\
Yu-Chen Chen$^1$, Chih-Yao Chen$^1$, Yu-Jung Lin$^1$, Yu-Ling Chen$^1$, Zih-Yu Chen$^1$, \\
I-Ning Tsai$^1$, Hsiu-Hsuan Wang$^1$, Ho-Lam Chung$^1$, Ke-Han Lu$^1$, Hung-yi Lee$^1$}}

\address{$^1$National Taiwan University $^2$University of Toronto}
\begin{document}
\ninept
\maketitle
\begin{abstract}
Large audio–language models are advancing rapidly, yet most evaluations emphasize speech or globally sourced sounds, overlooking culturally distinctive cues. This gap raises a critical question: can current models generalize to localized, non-semantic audio that communities instantly recognize but outsiders do not? To address this, we present TAU (Taiwan Audio Understanding), a benchmark of everyday Taiwanese “soundmarks.” TAU is built through a pipeline combining curated sources, human editing, and LLM-assisted question generation, producing 702 clips and 1,794 multiple-choice items that cannot be solved by transcripts alone. Experiments show that state-of-the-art LALMs, including Gemini 2.5 and Qwen2-Audio, perform far below local humans. TAU demonstrates the need for localized benchmarks to reveal cultural blind spots, guide more equitable multimodal evaluation, and ensure models serve communities beyond the global mainstream.

\end{abstract}

\section{Introduction}
\label{sec:intro}

Understanding the sounds around us often depends more on what is heard than on what is said. Everyday acoustic cues carry cultural meaning that is independent of language. Metro chimes, scooter beepers, and convenience-store jingles are recognized by locals because of exposure, not because of words. Contemporary audio evaluation already includes environmental sound \cite{audioset, fsd50k}. However, strong performance on many benchmarks can be achieved by matching generic categories that appear worldwide and are heavily represented in web-scale sources, such as siren, dog bark, doorbell, rainfall, and engine noise. This leaves open a key question: can current models recognize localized, non-semantic audio that communities instantly identify but that outsiders rarely encounter?

At the same time, \emph{Large Audio–Language Models (LALMs)}\cite{desta2.5_audio, qwen2_audio,Lu2025Developing,yang2024building} are recently becoming important backbones for multimodal assistants, accessibility tools, and embodied agents. 
Evaluation of modern LALMs has surged, but current benchmarks still emphasize speech semantics or generic, globally sourced environmental audio. 
LALM benchmarks, such as Dynamic-SUPERB \cite{huang2024dynamic,dynamicsuperb}, AIR-Bench\cite{airbench}, and MMAU\cite{mmau}, have advanced coverage across speech, paralinguistics, and some non-speech sounds; yet they were not designed to probe localized cultural knowledge in non-speech audio. They use other datasets sampled from global platforms (e.g., Freesound, YouTube). While this breadth is valuable, the taxonomies and sampling strategies seldom encode community-specific distinctiveness.

This gap carries significant implications, both for scientific methodology and societal equity.
Scientifically, failing to account for localized audio compromises the validity of evaluation benchmarks. 
Soundscape scholarship has long argued that communities possess characteristic “soundmarks”, distinctive auditory cues that anchor identity and place \cite{soundmark_urban}. 
Recognizing these depends on cultural exposure, not linguistic comprehension. 
Consequently, benchmarks that ignore such soundmarks create a distorted view of model performance, overestimating real-world competence while underestimating the generalization challenges across different geographic and cultural contexts.
These scientific oversights have direct societal consequences. 
Studies show that skewed or unrepresentative training data can perpetuate inequities and harm underrepresented communities. \cite{emo_bias, emo_debias, litsen_and_speak_fairly} 
Without deliberate localization, models trained on globally aggregated data will inevitably be deaf to community-specific signals. 
This leads to technologies that underperform for and marginalize populations outside the cultural and geographic mainstream.

We propose a pathway toward building localized audio benchmarks and present Taiwan as a case study. Our benchmark, TAU (Taiwan Audio Understanding) \footnote{https://dlion168.github.io/TAU\_demo/}, addresses this cultural localization gap. TAU curates everyday, locally distinctive Taiwanese non-speech sounds and evaluates models with multiple-choice questions (MCQs) that cannot be answered by semantic reasoning alone, steering evaluation toward timbre, rhythm, and iconic acoustic patterns. Beyond Taiwan, TAU illustrates how localized benchmarks can highlight cultural blind spots in audio–language models, include underrepresented communities in multimodal evaluation, and guide the design of more equitable and robust multimodal systems.

\begin{figure*}
    \centering
    \includegraphics[width=0.95\linewidth]{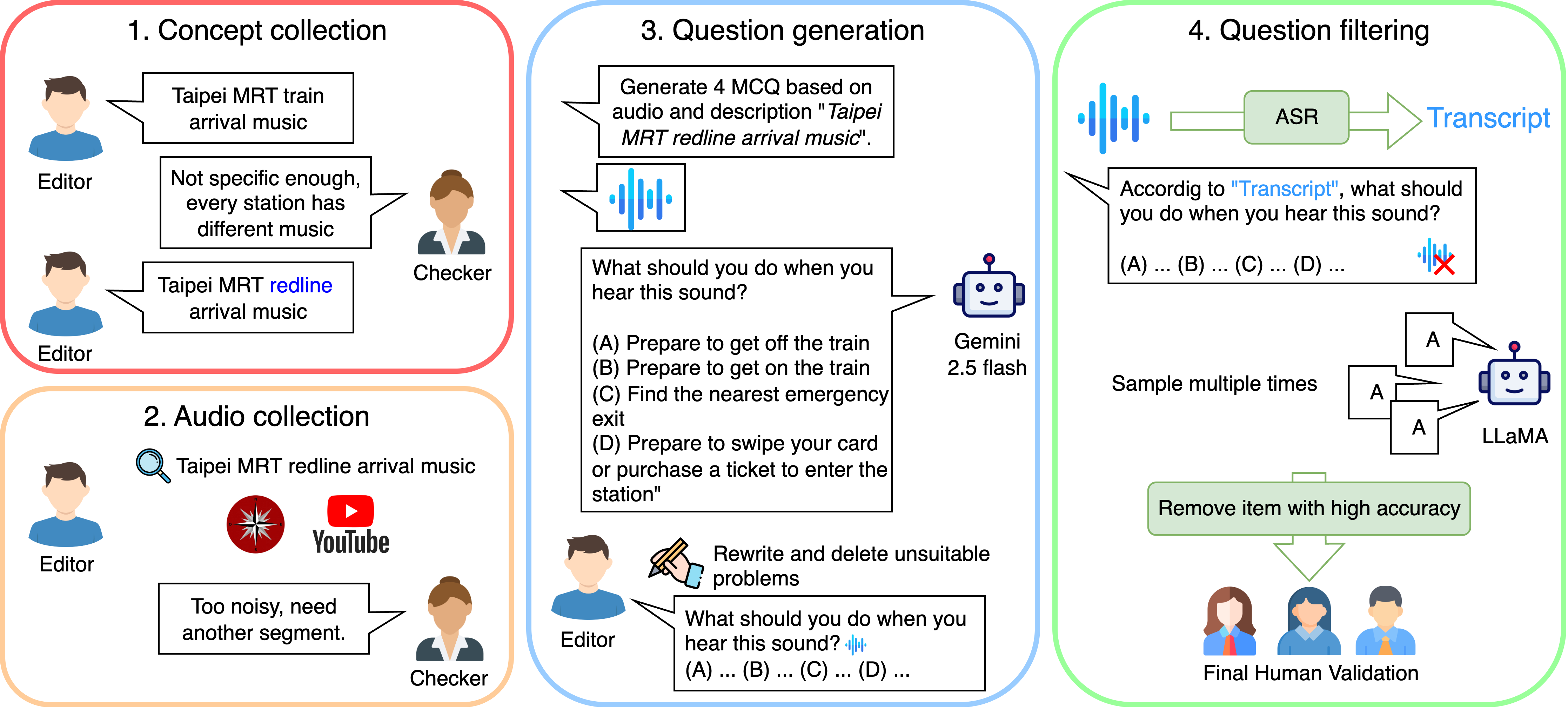} 
    \vspace{-4pt}
    \caption{Construction workflow of our localized audio benchmark TAU.}
    \label{fig:pipeline}
    \vspace{-7pt}
\end{figure*}






\section{Related works}

LALMs increasingly unify audio analysis with open-ended language outputs.  \textbf{Qwen2-Audio}\cite{qwen2_audio} reports strong instruction-following and audio analysis; \textbf{DeSTA2.5-Audio}\cite{desta2.5_audio} targets robust auditory perception and instruction following with a general-purpose LALM design, reporting broad improvements without task-specific audio instruction tuning; \textbf{Audio-Reasoner}\cite{audio_reasoner} focuses on long-context reasoning over auditory input; \textbf{Gemma-3n}\cite{gemma3} serves as a lightweight yet competitive model for audio–language tasks. Together, these LALMs illustrate the diversity of design choices and capabilities in the current landscape.  

In parallel, LALM evaluation has branched into three broad categories. (i) \textit{Chat-based audio understanding} benchmarks assess open-ended comprehension and dialogue grounded in audio, e.g. SD-Eval \cite{sdeval} for spoken dialogue beyond words, ADU-Bench \cite{adubench} for open-ended audio dialogue, and VoiceBench \cite{voice_bench} for LLM-based voice assistants. (ii) \textit{General-purpose multi-task audio understanding and reasoning} suites provide wide coverage across speech, non-speech, and paralinguistics, such as AudioBench \cite{audio_bench}, MMAU \cite{mmau}, and AIR-Bench \cite{airbench} for acoustic-aspect sensitivity. (iii) \textit{Domain-focused stress tests} target specific modalities or skills, including Audio Entailment \cite{audioentailment} for deductive reasoning, MuChoMusic \cite{muchomusic} for music understanding, SAKURA \cite{sakura} for audio muti-hop reasoning, ToxicTone \cite{toxictone} for toxic speech detection, and Speech-IFEval \cite{lu2025speechifeval} for instruction-following skills. While these efforts broaden coverage, most draw on globally sourced, generic corpora and rarely target locale-specific, non-semantic cues, which motivates our culturally localized TAU benchmark.

Recent benchmarks have begun probing country- or region-specific knowledge. BLEnD\cite{blend} evaluates locally grounded knowledge across 16 countries/regions and 13 languages, showing large culture-dependent gaps even for frontier models. CulturalBench \cite{culturalbench} provides 1,227 human-written, human-verified questions that span 45 global regions and 17 cultural topics. ThaiCLI \cite{thaicli} targets Thai cultural competence alongside core capabilities, illustrating how national-culture evaluation complements language proficiency assessments. TaiwanVQA\cite{taiwanvqa} and VisTW\cite{vistw} focus on Taiwanese culture through vision–language QA. However, these resources remain text- or image-centric. None evaluates whether models can recognize locale-specific, non-semantic acoustic cues that locals identify from sound alone. TAU complements culture-specific text or vision benchmarks by testing audio cultural grounding, a capability that current suites do not measure.
\vspace{-5pt}

\section{Dataset Collection}




\begin{figure*}
    \centering
    \includegraphics[width=1.0\linewidth]{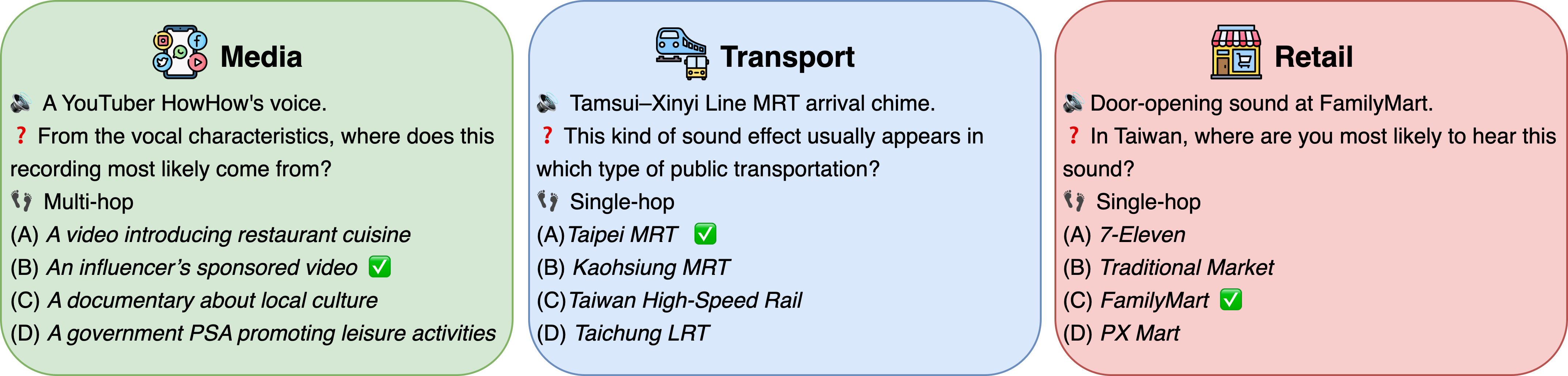} 
    \caption{Example TAU multiple-choice items in three categories \emph{Media}, \emph{Transit}, and \emph{Retail}, showing hop-type labels and culturally grounded distractors.}
    \label{fig:placeholder}
    \vspace{-4pt}
\end{figure*}
\subsection{Design Principles}
\label{sec:principles}
\textbf{Local identifiability} TAU targets sounds that Taiwanese listeners recognize immediately, while non-locals are unlikely to identify them without prior exposure. Curators prioritize everyday “soundmarks” (e.g., public signals, transportation cues, consumer-device prompts) rather than globally ubiquitous sounds.

\textbf{Semantic independence} Items must be answerable from the acoustic signature itself, using timbre, rhythm, envelope, or characteristic patterns, rather than lexical content. When speech fragments are present, they are trimmed or counter-balanced so that transcripts alone cannot reveal the answer.

\textbf{Diverse yet accessible} We encourage breadth across venues, devices, and activities, but avoid overly rare or expert-only phenomena. The aim is creative coverage of culturally salient sounds that typical residents encounter without specialized knowledge.

\subsection{Roles}
Annotators are 10 native Taiwanese recruited from different regions across Taiwan with a gender-balanced composition to ensure both regional and demographic diversity in practices and soundscapes. \emph{Editors} nominate culturally distinctive sounds and gather candidate audio sources. \emph{Checkers} verify that the proposed sounds meet the design principles in Sec.~\ref{sec:principles}, performing quality control on timing, audibility, and semantic independence. \emph{Editors} also refine question stems, options, and labels to ensure clarity and cultural plausibility. Finally, \emph{Reviewers} conduct human performance validation to confirm that items are both solvable and appropriately challenging.

\subsection{Workflow}
\label{sec:workflow}
The proposed dataset collection workflow is summarized in Fig.~\ref{fig:pipeline}.
\subsubsection{Concept collection}
Editors first compiled a pool of 550 Taiwan-specific “soundmarks,” each described with a short rationale for why it is instantly recognizable to locals but not to non-locals.
The collection emphasizes diversity across venues, devices, and activities while avoiding rare or expert-only edge cases. Checkers check if these sounds follow the design principles in Sec.~\ref{sec:principles}, producing a list of candidate sounds with provisional categories and concise descriptors.

\subsubsection{Audio Collection}
Approved targets are sourced from two channels: permissively licensed Creative Commons repositories (YouTube and aporee map) and self-recordings contributed by the team. Annotators provided metadata including \texttt{source url}, \texttt{start time}, and \texttt{end time}. To promote robustness beyond a single context, we attach up to three variants per target sound that differ in location, background conditions, or device/version (e.g., crowded vs. quiet carriage, different terminal stations). During scouting, segments can be up to 30\,s to facilitate selection. Each candidate must satisfy continuity (no abrupt cuts or dropouts), audibility of the intended source, and acceptable perceptual SNR as judged by human listeners. 

Every sourced clip is reviewed by a checker distinct from its proposer. The checker verifies timing format and boundary accuracy, confirms that the target sound dominates the excerpt, and performs a semantic-leakage screen to ensure the eventual item cannot be solved by lexical content alone. Descriptions are edited for specificity so that culturally grounded identity is unambiguous. For example, explicitly naming the soundscape from a specific metro line rather than simply “metro chime”. We collected 943 audios at this stage.

\subsubsection{Question Generation}
Quality-controlled clips are converted into four-option MCQs using a human-in-the-loop process. \emph{Gemini 2.5 Flash} automatically drafts four question stems, answer options, and tentative categories from minimal clip descriptors to avoid text-only shortcuts. Editors refine wording for clarity, calibrate distractors toward plausible near-miss cultural confusions, delete unsuitable items, and diversify what each item probes so that the four questions attached to a clip assess different facets (e.g., place, source object, activity, cultural practice) rather than repeating a single skill. Each item is labeled as \emph{Single-hop} when a single acoustic cue suffices, or \emph{Multi-hop} when the answer requires combining acoustic evidence with background knowledge.

\begin{figure}[tbp!]
  \centering
  \begin{tikzpicture}
    \begin{axis}[
      ybar,
      bar width=7pt,
      width=8cm,
      height=4.5cm,
      ylabel={Count},
      ymin=0,
      ymax=440,
      symbolic x coords={Retail,Cultural,Announcement,Education,Transit,Media,Entertainment,Nature,Emergency,Payment},
      xtick=data,
      xticklabel style={rotate=45, anchor=east},
      nodes near coords,
      enlarge x limits=0.05,
    ]
      \addplot coordinates {
        (Retail,69)
        (Cultural,261)
        (Announcement,149)
        (Education,104)
        (Transit,241)
        (Media,288)
        (Entertainment,385)
        (Nature,107)
        (Emergency,36)
        (Payment,154)
      };
    \end{axis}
  \end{tikzpicture}
  \vspace{-10pt}
  \caption{Distribution of question types in TAU benchmark.}
  \label{fig:type_bar}
\end{figure}
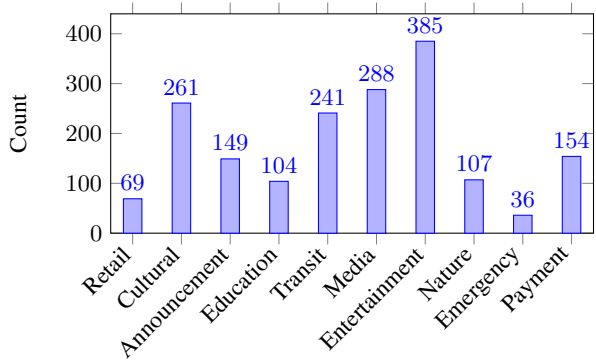

\vspace{-8pt}
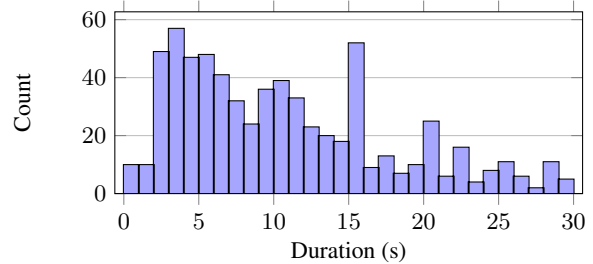
\begin{figure}[tbp]
  \centering
  \begin{tikzpicture}
    \begin{axis}[
      ybar,
      bar width=6pt,
      width=7.8cm,
      height=4.0cm,
      xlabel={Duration (s)},
      ylabel={Count},
      ymin=0,
      xmin=0, xmax=30,
      xtick={0,5,10,15,20,25,30},
      ymajorgrids,
      enlarge x limits=0.02
      ]
    ]
      \addplot+[draw=black, fill=blue!35] coordinates {
        (0.5,10)
        (1.5,10)
        (2.5,49)
        (3.5,57)
        (4.5,47)
        (5.5,48)
        (6.5,41)
        (7.5,32)
        (8.5,24)
        (9.5,36)
        (10.5,39)
        (11.5,33)
        (12.5,23)
        (13.5,20)
        (14.5,18)
        (15.5,52)
        (16.5,9)
        (17.5,13)
        (18.5,7)
        (19.5,10)
        (20.5,25)
        (21.5,6)
        (22.5,16)
        (23.5,4)
        (24.5,8)
        (25.5,11)
        (26.5,6)
        (27.5,2)
        (28.5,11)
        (29.5,5)
      };
    \end{axis}
  \end{tikzpicture}
  \vspace{-8pt}
  \caption{Histogram of audio durations in TAU benchmark.}
  \label{fig:duration_hist_cont}
\end{figure}

\subsubsection{Question filtering}
After editor revisions, we employ an automatic item filtering pipeline. A strong ASR model (Whisper large v3 \cite{whisper}) first transcribes any spoken content in the clip. Then, a text-only LLM (LLaMA-3.1 8B \cite{llama3.1}) attempts to answer the multiple-choice questions using only the transcript, without access to the audio. This procedure identifies items that can be solved through lexical cues alone. For each item, we sampled five responses and computed the model’s success rate. We then perform a one-tailed $t$-test under the null hypothesis $\mathcal{H}_0$: the model’s success rate does not exceed random guessing (25\%). If the null is rejected at $p < 0.05$, the item is considered vulnerable to transcript-only shortcuts and is discarded. This ensures that the retained questions genuinely require acoustic or cultural understanding rather than textual leakage. 

\subsection{Dataset Statistic}
Fig.~\ref{fig:type_bar} and ~\ref{fig:duration_hist_cont} summarize the scale and composition of TAU. 
It contains 702 audio clips across 10 culturally distinctive categories. 
This diversity ensures that the dataset covers both highly frequent urban soundmarks (e.g., transit chimes, store jingles) and less common but socially important cues (e.g., emergency alarms, religious chants). 
The category distribution is intentionally imbalanced to reflect the natural frequency of sounds in everyday Taiwanese soundscapes, rather than enforcing artificial uniformity. 

Each clip is paired with up to four MCQs, resulting in 1794 evaluation items in total. 
The median clip length is 9.43 seconds, with a maximum of 30 seconds by design. This design balances realism with usability, allowing evaluation without excessive cognitive load.
On average, each soundmark has 2.1 recording variants that differ by location, time, or background conditions, which increases robustness and reduces overfitting to specific contexts.


\begin{table}[tbp]
\centering
\caption{Model performance on TAU benchmark using default system prompt, separated by Single-hop and Multi-hop items (Accuracy \%). Parameter counts (Params) are reported in billions. Best results in \textbf{bold}.}
\setlength{\tabcolsep}{4pt}
\footnotesize
\label{tab:performance}
\vspace{2mm}
\begin{tabular}{l l r c c}
\toprule
\textbf{Category} & \textbf{Model} & \textbf{Params} & \textbf{Single-hop} & \textbf{Multi-hop} \\
\midrule
Random & Random & -- & 25.0 & 25.0 \\
\midrule
\multirow{8}{*}{LALM} 
 & Gemini 2.5 Pro       & --   & 72.4 & 73.9 \\
 & Gemini 2.5 Flash     & --   & 61.3 & 63.2 \\
 & Qwen2-Audio-Instruct & 8.2B & 30.3 & 27.8 \\
 & Qwen2.5-Omni-7B      & 7.6B & 46.4 & 46.1 \\
 & DeSTA2.5-Audio       & 8.8B & 43.3 & 41.7 \\
 & Gemma-3n-E2B-it      & 4.4B & 29.6 & 25.8 \\
 & Gemma-3n-E4B-it      & 6.8B & 29.0 & 25.9 \\
\midrule
ASR+LLM & LLaMA-3.1      & 9.6B & 34.9 & 34.1 \\
\midrule
\multirow{2}{*}{LLM only} 
 & Qwen2.5-7B-Instruct  & 7.0B & 38.5 & 35.5 \\
 & LLaMA-3.1            & 8.0B & 37.6 & 41.4 \\
\midrule
Human & Topline & -- & \textbf{84.0} & \textbf{83.3} \\
\bottomrule
\end{tabular}
\vspace{-8pt}
\end{table}

\vspace{-10pt}
\section{Experiments}
\vspace{-2pt}
\subsection{Evaluation protocol}
We evaluate model performance on TAU using several LALMs. Each model is tested under two system prompts: (i) its default prompt, and (ii) a culturally grounded prompt, \textit{“You are a Taiwanese person. Always respond with the perspective, cultural background, and knowledge of someone from Taiwan.”} We parse the output of all models to 4 options via Gemini-2.0 flash. For comparison, we include several baselines: \textit{Random}, which selects an answer uniformly at random; \textit{ASR+LLM}, which feeds Whisper-large-v3 transcriptions of the clips to a text-only LLM; and \textit{LLM only}, which asks the same questions directly without any audio or transcript input. Finally, we report a \textit{Human} topline: nine annotators each answer the benchmark, with every question evaluated by two independent annotators.

\begin{table}[tbp]
\centering
\caption{Model performance on TAU benchmark using culturally grounded prompt (Accuracy \%). Best results in \textbf{bold}.}
\vspace{2mm}
\footnotesize
\label{tab:performance_cultural}
\begin{tabular}{l l c c}
\toprule
\textbf{Category} & \textbf{Model} & \textbf{Single-hop} & \textbf{Multi-hop} \\
\midrule
Random & Random & 25.0 & 25.0 \\
\midrule
\multirow{8}{*}{LALM} 
 & Gemini 2.5 Pro       & \textbf{70.6} & \textbf{71.8} \\
 & Gemini 2.5 Flash     & 62.8 & 62.2 \\
 & Qwen2-Audio-Instruct & 29.0 & 27.1 \\
 & Qwen2.5-Omni-7B      & 43.6 & 42.3 \\
 & DeSTA2.5-Audio       & 38.2 & 38.9 \\
 & Gemma-3n-E2B-it      & 29.7 & 29.4 \\
 & Gemma-3n-E4B-it      & 34.0 & 33.4 \\
\midrule
ASR+LLM & LLaMA-3.1 & 34.7 & 31.8 \\
\midrule
\multirow{2}{*}{LLM only} 
 & Qwen2.5-7B-Instruct   & 38.4 & 34.3 \\
 & LLaMA-3.1 & 37.7 & 35.8 \\
\bottomrule
\end{tabular}
\vspace{-8pt}
\end{table}

\subsection{Performance with Default System Prompt}
Table~\ref{tab:performance} presents results with each model’s default prompt. 
Overall, humans remain far ahead of all systems, achieving 84.0\% on Single-hop and 83.3\% on Multi-hop questions. 
Among LALMs, \textbf{Gemini 2.5 Pro} clearly leads (72.4\% / 73.9\%), followed by \textbf{Gemini 2.5 Flash} (61.3\% / 63.2\%). 
Other audio–language models such as Qwen2.5-Omni-7B, DeSTA2.5-Audio, and Qwen2-Audio-Instruct show moderate performance (30–46\%), while lightweight models such as Gemma-3n variants struggle to surpass random baselines by a large margin.  

Interestingly, text-only baselines reveal limitations of transcript reasoning. 
The \textit{ASR+LLM} setup with LLaMA-3.1 achieves 34.9\% / 34.1\%, and \textit{LLM only} baselines (Qwen-2.5, LLaMA-3.1) stay around 35–41\%. 
This confirms that most items cannot be solved by lexical information alone, validating the design of TAU.  
We also observe that Multi-hop accuracy lags behind Single-hop across weaker systems, suggesting that models find it harder to integrate cultural background knowledge with acoustic evidence.  
\vspace{-8pt}
\subsection{Performance with Local Specific Prompt}

We next examine results under the culturally grounded prompt. 
Table~\ref{tab:performance_cultural} shows that this prompt does not universally improve performance, but produces nuanced shifts.  

For high-performing models such as \textbf{Gemini 2.5 Pro}, accuracy slightly decreases compared to the default setting (70.6\% / 71.8\% vs. 72.4\% / 73.9\%), suggesting that localization cues may not help frontier systems already capable of strong acoustic reasoning.  
By contrast, \textbf{Gemma-3n} variants benefit more noticeably, improving from $\sim$29\% to 33–34\% on Multi-hop items, narrowing the gap with mid-tier LALMs.  
Some models, such as DeSTA2.5-Audio and Qwen2.5-Omni-7B, show mixed changes, indicating that cultural priming can alter reasoning strategies but does not consistently enhance accuracy.  

Overall, the culturally specific prompt highlights that \emph{prompting alone is insufficient} for closing the gap between models and humans. 
Nevertheless, its selective gains for certain models suggest a pathway for aligning LALMs with localized knowledge, especially for weaker or instruction-tuned models. 
Future work could explore integrating explicit cultural grounding into training rather than relying solely on prompt engineering.

\vspace{-8pt}
\section{Limitations}
\vspace{-6pt}
\label{sec:limitations}
First, TAU is intentionally centered on Taiwan-specific soundmarks. Strong performance on TAU does not guarantee competence in other locales; weak performance may reflect cultural unfamiliarity rather than general auditory shortcomings. Secondly, our pool of sounds, venues, devices, and times of day is finite. Urban scenes may be overrepresented relative to rural contexts, and some cultural practices may be undersampled. Variation in devices and microphones also introduces covariate shifts that are not fully controlled. Lastly, soundmarks change over time (e.g., updated transit chimes or jingles). Models evaluated today may face a distribution shift as the physical environment evolves. We would adopt a “versioned benchmark” philosophy, release snapshots with explicit collection dates, and encourage lightweight incremental updates rather than one-off frozen datasets.
\vspace{-8pt}
\section{Conclusion}
\vspace{-6pt}
We presented a pathway for constructing \emph{localized} audio benchmarks and instantiated it with a culturally grounded, non-semantic evaluation suite. Our five-stage pipeline, covering concept curation, licensed sourcing, quality control, LLM-assisted item generation, and leakage filtering, offers a reproducible template that others can adapt to different regions and communities. Empirically, current audio–language models fall well short of human performance on localized items, and a culturally specific prompt provides, at best, selective gains, indicating that prompt engineering alone is insufficient. These findings underscore the need for culturally informed data, tasks, and metrics to assess real-world robustness, mitigate region-dependent failure modes, and support equitable deployment of multimodal systems.
\bibliographystyle{IEEEbib}
\bibliography{strings,refs}

\end{document}